# Neuroblastoma: nutritional strategies as supportive care in pediatric oncology

Hafida Hamdache[1] ✉ , Alexia Gazeu[2,3] , Marion Gambart[4] , Nathalie Bendriss-Vermare[2] , Vera Pancaldi[1]✉

Author Affiliations:

1. CRCT, Université de Toulouse, Inserm, CNRS, Toulouse III-Paul Sabatier University, Cancer Research Center of Toulouse, France.
2. CISTAR team, Cancer Research Center of Lyon, INSERM U1052, CNRS UMR5286, Université de Lyon, Université Lyon 1, Centre Léon Bérard, F-69000 Lyon, France
3. Department of Pathology, Hôpital Femme-Mère-Enfant, Hospices Civils de Lyon, University Hospital of Lyon, 59 Bd Pinel, 69500, Lyon Bron, France.
4. Pediatric Oncology and Hematology Department, Toulouse University Hospital, Toulouse, France

Corresponding authors:

✉ Hafida Hamdache hafida.hamdache@inserm.fr ; Phone number : +33582741693

✉ Vera Pancaldi vera.pancaldi@inserm.fr ; Phone number : +33582741693

## Abstract

Neuroblastoma, is a highly heterogeneous pediatric tumour and is responsible for 15% of pediatric cancer-related deaths. The clinical outcomes can vary from spontaneous regression to high metastatic disease. This extracranial tumour arises from a neural crest-derived cell and can harbor different phenotypes. Its heterogeneity may result from variations in differentiation states influenced by genetic and epigenetic factors and individual patient characteristics. This leads downstream to disruption of homeostasis and a metabolic shift in response to the tumour's needs. Nutrition can play a key role in influencing various aspects of a tumour's behaviour. This review provides an in-depth exploration of neuroblastoma's aetiology and the different avenues of disease progression, which can be targeted with individualized nutrition intervention strategies to improve the well-being of children and optimize clinical outcomes.

# 1. Introduction

Neuroblastoma (NB) is the second most common extracranial solid tumour of childhood and it accounts for 15% of pediatric tumour deaths. NB exhibits diverse clinical, histological, and biological characteristics, leading to a wide variety of outcomes. It can range from tumours that regress spontaneously to highly aggressive and metastatic cases, with 5-year survival rates ranging from 90% to 95% for low- (LR) and intermediate-risk (IR) disease, to 40%–50% for high-risk (HR) disease.[1]

NB arises from neural crest-derived sympathoadrenal cells during embryonic development. The average age at diagnosis is 18 months, and in 90% of cases, NB occurs before the age of 5 years.[2] Primary tumors predominantly occur in the adrenal medulla (50-60%) and sympathetic ganglia (thorax 20%, neck 5%, pelvis 5%). At diagnosis, 50% of cases present with metastases, commonly found in the bone marrow, cortical bones, lymph nodes, liver, and brain.[3] Clinical symptoms depend on the tumour's location and extent. Common presentations include abdominal mass (primarily in the adrenal medulla), Horner syndrome (disrupted facial sympathetic innervation), airway compromise (with neck or thoracic tumours) and spinal cord or cauda equina compression (invasion of neural foramina). Infants may present with extensive liver disease, leading to synthetic dysfunction, coagulopathy, or abdominal compartment syndrome. Metastatic disease can manifest with fever, malaise, pain, and cytopenias.[4] NBs exhibit one of the highest spontaneous regression rates among all cancers, particularly notable in the LR subsets, possibly through immune-mediated elimination.[1,5] Patients with Opsoclonus-Myoclonus syndrome (OMS), a rare disorder that arises as an indirect effect of cancer and predominantly found in LR NB cases with neurological symptoms (involuntary eye movements and muscle spasms) often show immune cells infiltration within the tumoural tissues. These cells will cluster and form tertiary lymphoid structures (TLS), which may be induced by an unclear immune reaction against NB cells.[6]

Molecular profiling has allowed improvements in HR group stratification. However, it is crucial to note that the tumour microenvironment (TME) is a complex dynamic system. It includes an extracellular matrix, different cell types that communicate and perform diverse functions establishing a specific metabolic state for the cancer cells. For immune cells to be effective, they require appropriate metabolism.

Like in all other solid tumours, NBs cells hijack the nutrients available in the TME for cell growth and proliferation, at the disadvantage of immune cells.[7,8] Furthermore, the activation of immune cells also depends on the patient's history, including previous infections, antibiotic use, diet, and microbiota diversity.[9] Recent findings indicate that the microbiome of NB patients was altered compared to healthy individuals and their siblings.[10] Breastfeeding and maternal use of supplements containing folic acid, vitamins, or minerals during the preconception period were inversely associated with NB. Plasma metabolite analysis in NB patients also demonstrated that these metabolites could be used to correctly classify patients by age, risk level, and response to therapy, but not MYCN status.[11]

These findings bring to light the importance of considering the molecular and metabolic states of the tumour, as well as the overall health and history of the patient, in designing effective treatment strategies. This review provides a comprehensive description of NB's initiation and progression, highlighting the roles of immune deficiency, gut microbiome alterations, and metabolic shift. Crucially, it offers nutritional strategies as promising supportive care options for clinicians, enhancing the effectiveness of current therapies.

## 2. Neuroblastoma: a highly heterogeneous disease

### 2.1 Neuroblastoma lineage drives tumour heterogeneity

Initially, it was thought that NB originated from NCCs committing to the adrenergic lineage and differentiating into chromaffin cells, neuroendocrine cells found mostly in the medulla of the adrenal glands. The adrenergic phenotype refers to this type of cell, capable of producing catecholamines such as adrenaline. However, it has been demonstrated that chromaffin cells can arise from two distinct lineages: sympathoadrenal progenitor cells and Schwann cell precursors (SCPs) that can differentiate into many types of cells like enteric neurons, melanocytes, odontoblasts and Schwann cells that produce myelin protecting and isolating nerves fibers.[12,13] (Figure 1). SCPs also differentiate into sympathoblasts, which subsequently develop into chromaffin cells through a transition stage known as bridge cells.[13] This highlights the considerable plasticity of SCPs and emphasizes the need to consider both lineages in understanding the origin and heterogeneity of NB. Furthermore, NB cells can have a spontaneous and reversible phenotype between mesenchymal and adrenergic, associated with epigenetic reprogramming. The mesenchymal phenotype is characterized by increased motility and invasiveness and is often linked to stem-like traits.[14] The degree of differentiation in NB cells may vary based on accumulation and types of mutations that affect specific biological pathways, the epigenome and transcriptional programmes influencing cell fate. This variety of phenotypes, underlined by differentiation stages, could potentially explain the observed diversity in patients' survival.

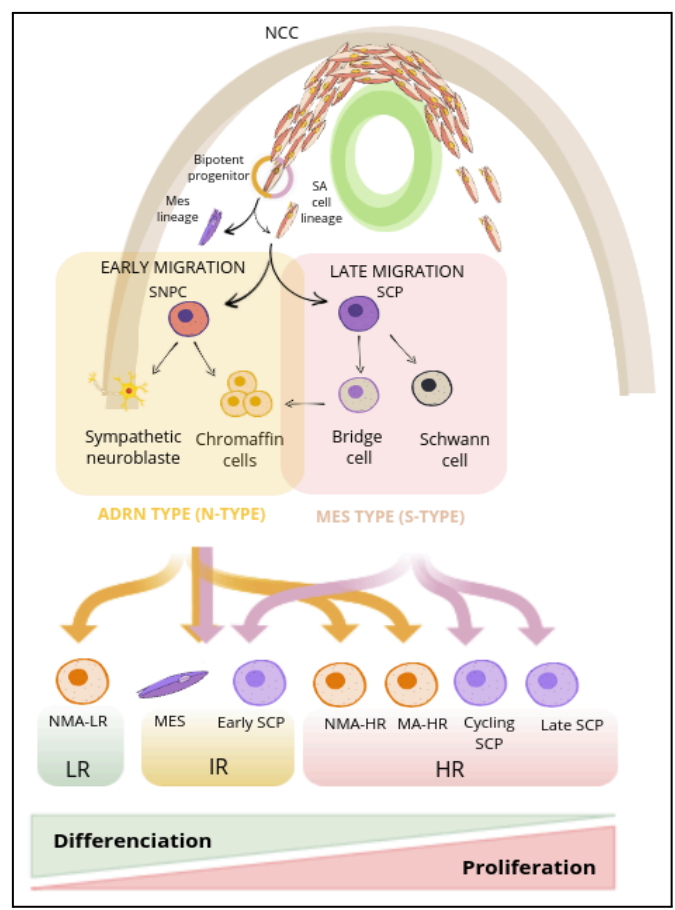

*Figure* 1: Developmental Pathways and Cellular Differentiation in Neuroblastoma Subtypes[13]
Differentiation trajectories of neuroblastoma cells originating from neural crest cells (NCC) illustrate distinct scenarios for adrenergic (ADRN) and mesenchymal (MES) subtypes, emphasizing the correlation between cellular differentiation stages and neuroblastoma risk categories: low risk (LR), intermediate risk (IR), and high risk (HR). LR includes non-MYC amplified cells (NMA-LR). IR comprises mesenchymal cells (MES) and early Schwann cell precursors (early SCPs). Finally, HR consists of NMA-HR, cycling cells, and late SCPs.

NB's heterogeneity was attributed to the identification of two core regulatory circuits leading to two distinct populations of tumour cells: a more mesenchymal NCC-like cell type, which lacks noradrenergic markers and is characterized by the expression of transcription factors (TFs) in the FOS and JUN families; and a more committed noradrenergic cell, marked by enzymes like tyrosine hydroxylase and dopamine beta hydroxylation and involving expression of PHOX2B, HAND2 and GATA3 TFs.[15,16] Recent studies confirmed these results and are in alignment with our hypothesis: a single-cell transcriptomic analysis allowed the identification of three distinct tumour groups.[17] Group A, characterized by undifferentiated chromaffin-like cells phenotype and associated with HR NB, includes most MYCN amplified tumours with an enhanced epithelial mesenchymal transition NCCs phenotype. Group C, presented a highly differentiated chromaffin-like cell signature, associated with LR NB. Finally, group B displayed both undifferentiated and differentiated chromaffin-like cell signatures, likely representing mixed tumours with the two contingents. Another single-cell study on human adrenal glands at various developmental stages and NB tumours, reinforcing that different NB risk groups correlate with distinct neuroblast developmental trajectories. LR NB tumours were more aligned with differentiated neuroblasts whereas HR NB with mesenchymal features were more closely mapping to the crossroad of early neuroblast lineage and chromaffin. [18,19]

## 2.2 Genetic and epigenomic alterations affecting cellular phenotypes and plasticity in NB

NB derives from a neural crest cell (NCC), which has an inherent plasticity and an undifferentiated phenotype necessary for cell fate determination and normal development. NB is likely associated with the lack of acquisition of a well-differentiated phenotype during embryonic development influenced by genetic and epigenetic factors.

NB initiation is triggered by germline genetic alterations in 2% of all cases, with familial NB caused by germline genetic alterations of paired-like homeobox 2B (PHOX2B) and ALK (Anaplastic lymphoma kinase) (Table. 1).[20] Both of these genes play an important role in the acquisition of a sympathetic nervous system lineage. NB can exhibit chromosomal alterations, the most commonly known include deletion of chromosome arms 1p (1p-del), 11q (11q-del), 19p, or gain of 17q (17q-gain) (Table. 1).[21–23] These alterations lead to dysfunctional signaling or chromatin remodeling during embryonic development, enhancing tumour initiation, promoting growth and inducing a less differentiated tumour phenotype, which contributes to more aggressive forms of NB.[24–28]

NB also harbors a wide spectrum of somatic mutations in several genes. The gene encoding the transcription factor MYCN is the most famous one and is correlated with a poor prognosis. [3,29] MYCN amplification during normal development is necessary for coordinating migration and NCCs fate decisions by regulating specific genes involved in the cell cycle, proliferation, and the maintenance of a pluripotency state. MYCN expression gradually decreases over time allowing the cell's differentiation.[30] In pathologic conditions, the high level of MYCN expression is maintained through multiple mechanisms, including genetic amplification, transcriptional activation, reduced degradation via the proteasome and increased stabilization of the protein.[31]

Thus, MYCN helps NB cells maintain their undifferentiated phenotype, enhancing their plasticity and adaptive capacity. MYCN is also cooperating with other mutated genes like ALK, which is also subject to somatic mutation disrupting the differentiation of NCCs within the sympathetic ganglia.[32] Overexpression of the LIN28B (Lin-28 homolog B) gene is correlated with poorer clinical outcomes of NB patients, via its role in preventing differentiation.[33] Protein tyrosine phosphatase 11 (PTPN11) is also found to be mutated in patients with worst clinical outcomes.[34] PTPN11 promotes the activation of the RAS/MAPK pathway and high RNA level expression is associated with worst survival in NB.[4] Finally, ATRX (alpha thalassemia/mental retardation syndrome X-linked) alterations induce a shift from silencing chromatin regions to active gene promoters, through a repressive complex that enhances the silencing of neuronal differentiation genes.[4,35] (Table. 1)

| Genetic alteration | Gene | Risk Group | Frequency | Biological impact |
| --- | --- | --- | --- | --- |
| Mutation | PHOX2B | Familial NB | ~2% | Differenciation |
| Mutation / Amplification | ALK | Familial NB<br>Somatic mutations | ~2%<br>~10% | Signaling dysregulation |
| Chromosomal alteration | 1p-del | IR/HR | 30% | Chromatin remodeling and epigenetic regulation |
| Chromosomal alteration | 11q-del | All risk groups | 35~45% | DNA repair and genomic stability |
| Chromosomal alteration | 19p-del | All risk groups | Correlated with older patients | Potentially altering cell communication |
| Chromosomal alteration | 17q-gain | All risk groups | 35~45% | Apoptosis and immune evasion |
| Mutation / Amplification | MYCN | HR | 25% | Metabolic reprogramming |
| Mutation | LIN28B | HR | ~10-15% des cas | Differentiation and metabolism |
| Mutation | PTPN11 | All risk groups | 3·4% | Signaling dysregulation |
| Mutation | ATRX | HR | ~10% | Chromatin remodelling and epigenetic regulation |

*Table* 1: **Summary of genetic alterations associated with neuroblastoma** [3,4,29,34,36]

## 2.3 The non-immune TME is a potential determinant of cell fate in NB

The surrounding local tissue also seems to play a key role in NB tumorigenesis. In physiological conditions, NCCs migration, delamination and lineage engagement are regulated by epigenetic and transcriptional programs influenced by bone morphogenetic protein (BPM), Wnt and FGF signaling.[13] Signals from sympathetic ganglia and especially olfactomedin-1 drive NB cells to shift from a noradrenergic into a more mesenchymal phenotype and activate a metastatic program.[37] Discoidin domain receptor tyrosine kinase 2, a mechanosensing tyrosine kinase receptor involved in many biological functions, was found to be downregulated in NB cells and induced transcriptional changes leading to reduced proliferation and a shift toward a more senescent phenotype [38]. Additionally, Wnt signaling was found to play a key role in NB cell lineage determination. Wnt3a/Rspo2 treatment on NB cells induced an epithelial-to-mesenchymal transition and neural differentiation, with increased activity of key TFs like SNAIL1/2 and TWIST1/2.[39] RNA sequencing has linked specific Wnt-responsive gene subsets to different prognostic outcomes in NB. These studies emphasize the TME's influence on NB cell fate, highlighting its impact on NB initiation and progression. The scenario of genetic alterations, chromosomal arrangements, transcriptomic and epigenetic changes in NB may contribute to increased tumour aggressiveness by the disruption of cellular homeostasis and tumour cell metabolism.

# 3. Genetic alterations are associated with metabolic profiles

In order to proliferate, NB cells adjust their metabolic fluxes to provide all the necessary elements for cell growth and division. This includes lipids to create cell membranes, storing and producing energy to ensure cellular processes and acting as signaling molecules, amino acids as the building blocks for protein synthesis, and nucleotides for DNA and RNA synthesis. Studies performed on NB cells metabolism highlight their metabolic adaptability associated by transcriptomic changes.

### 3.1 Energy metabolism

Glycolysis. NB cells preferentially increase glucose uptake and convert it into lactate (Warburg effect), even in normoxic conditions. This shift allows ATP production and supports tumour growth. MYCN TF induces a metabolic reprogramming in NB by upregulating key glycolytic enzymes like hexokinase 2 (HK2), pyruvate dehydrogenase kinase 1 (PDK1) and lactate dehydrogenase A (LDHA) (Figure 2).[29] In hypoxic conditions, HIF-1α further enhances glycolysis by binding to hypoxia response elements in these gene promoters. This synergistic effect of MYCN and HIF-1α ensures that NB cells can sustain high levels of glycolytic activity in response to their increased metabolic needs.

Tricarboxylic acid (TCA) cycle and oxidative phosphorylation. The TCA cycle occurs in the mitochondria, it supervises coenzymes production that are essential for energy production and maintains a pseudo homeostasis in the redox balance which might already be disrupted to cope with the high proliferation. It produces coenzymes like NADH and $FADH_2$ by oxidizing acetyl-CoA from carbohydrates, fats, and proteins. These coenzymes are crucial for ATP production through oxidative phosphorylation and help regulate the $NADH/NAD^+$ and $FADH_2/FAD$ ratios, supporting various biochemical reactions and cellular metabolism. Several enzymes involved in TCA cycle like Citrate Synthase (CS), Isocitrate Dehydrogenase Isoform 2 (IDH2) and Alpha-Ketoglutarate Dehydrogenase-Like (OGDHL) were overexpressed in MYCN amplified NB cells and tumours samples (Figure 2).[40] Furthermore, subunits of the respiratory chain and ATP synthase were also upregulated. Finally, NB patients with the worst survival were associated with overexpression of enzymes of the TCA cycle. MYCN amplified NB cells had higher oxygen consumption rates reflecting a higher oxidative phosphorylation activity compared to noMYCNN amplified NB cells.

Fatty acid and β oxidation. β oxidation also occurs in the mitochondria and is responsible for lipid degradation providing cells with other energy fuel sources. Lipid catabolism leads to acetyl CoA production that can either be used in the TCA cycle or released in the cytoplasm via the carnitine ship. Some specific enzymes involved in β oxidation, like hydroxyacyl-CoA dehydrogenase (HADH), were expressed at higher levels in MYCN amplified NB cells (Figure 2) . This increase was associated with poor survival in NB patients[29]. Oxygen consumption rate was found to be increased with the presence of palmitate and both basal and maximum consumption rates were higher in MYCN amplified NB cells compared to MYCN non-amplified NB cells. Finally, metabolic fuel is also ensured by NB-derived extracellular vesicles that carry lipids that will be catabolized through β oxidation. Furthermore, these vesicles also can carry metabolic enzymes involved in glycolysis, TCA cycle, lipogenesis, electron transport chain and glutaminolysis allowing a functional metabolic network within the TME.[41]

### 3.2 Amino acid metabolism

Several studies have highlighted the role of some specific amino acids in NB metabolism particularly in HR NB.

Glutamine. MYCN amplified NB cells are able to increase their glutamine uptake and synthesis in order to support tumour growth, survival and metabolic adaptation in the TME. The ASCT2 glutamine receptor's expression is upregulated by MYCN and ATF4, particularly under stress conditions (Figure 2).[29,40,42] MYCN is also able to upregulate several enzymes involved in glutaminolysis like glutaminase 2 (GLS2) which transforms glutamine into glutamate[29]. Glutamate is then converted into α-ketoglutarate (α KG) via glutamate dehydrogenases (GLUD1/2) and transaminases. These enzyme expressions are enhanced by MYCN-ATF4 -Lysine Demethylase 4C (KDM4C) cooperation.[29]

Cysteine. Cysteine is a non-essential amino acid that plays a role in metabolic reprogramming in cancer.[43] Under stress conditions it is involved in the production of glutathione, thus acting on the control of the redox balance. It is also involved in energy production through hydrogen sulfide (H2S) which boosts mitochondrial activity and energy production. Finally, cysteine is a carbon source for cancer cells to grow and regulate genes and it can be uptaken from the TME via excitatory amino acid transporter 3 (EAAT3) and the alanine-serine-cysteine-transporter 2 (ASCT2) or the cysteine–glutamate antiporter transport system (xCT).[43] In NB cells xCT is transcriptionally upregulated by MYCN. Furthermore, KDM4C and ATF4 induce the upregulation of SLC3A2 and SLC7A11 subunits of xCT as well as cystathionine- β synthase (CBS) and cystathionase (CTH) two enzymes involved in the transsulfuration pathway (Figure 2). CBS catalyzes the condensation of homocysteine with serine to form cystathionine, which is then hydrolyzed by CTH to generate cysteine and α-KG.[43]

*Serine glycine pathway.* This pathway takes its root from the glycolysis, 3 phosphoglycerate (3PG) is turned successively into 3 phosphohydroxypyruvate (3PHP), 3 phosphoserine, serine and finally glycine.[29] These metabolic reactions are catalyzed by phosphoglycerate dehydrogenase (PHGDH), phosphoserine aminotransferase 1 (PSAT1), phosphoserine phosphatase (PSPH) and finally serine hydroxymethyltransferase 1/2 (SHMT1/2) (Figure 2). Alongside α-KG and NADH are produced feeding the TCA cycle and respiratory mitochondrial chain and 5,10-methylenetetrahydrofolate (5,10-MTHF) that is used in several metabolic pathways among them one carbon metabolism leading to nucleotides production and amino acid methylation.[29] NB tumours that are MYCN amplified showed an increase in serine and glycine levels. Up-regulation of PHGDH, PSAT1, SHMT2 were observed and induced by MYCN-ATF4 that bind to their respective promoters.[29]

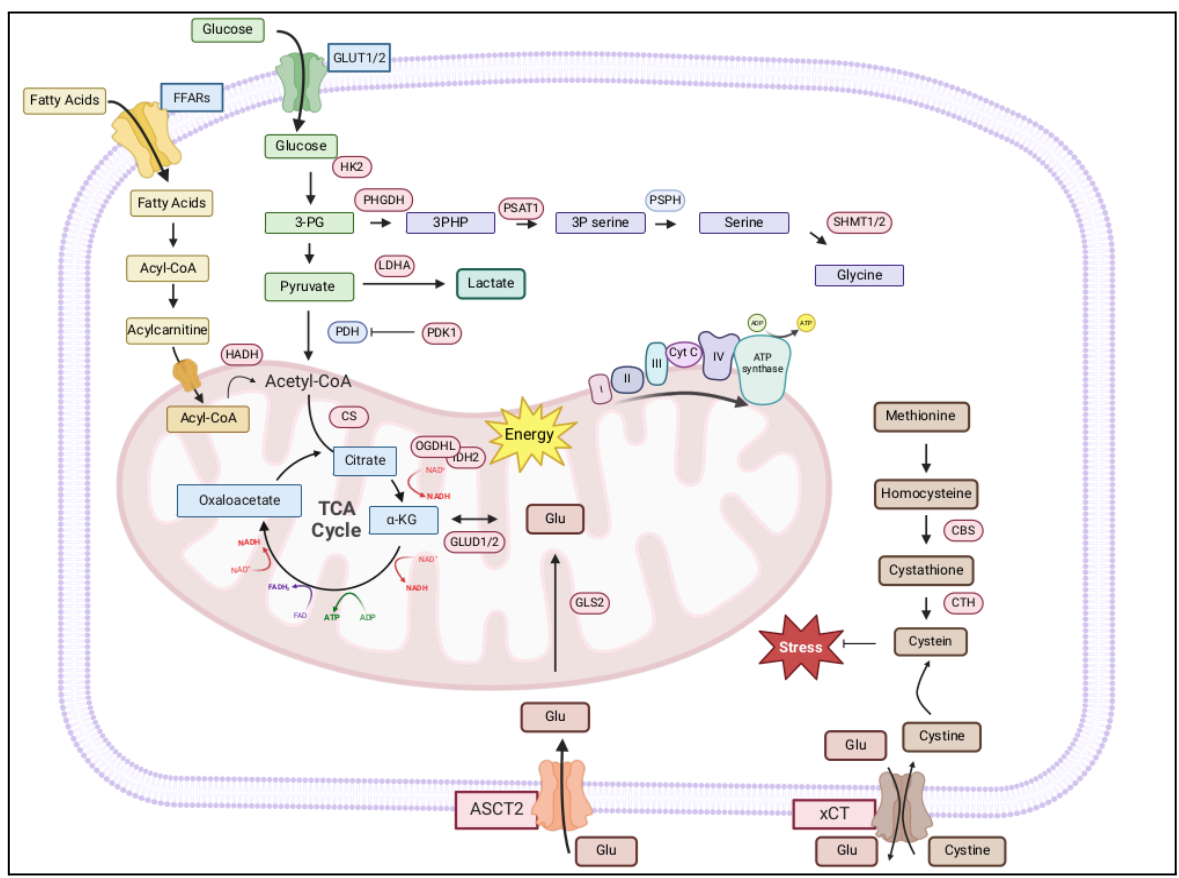

*Figure* 2: **Metabolic adaptation in NB induced by genetic alteration to promote cell proliferation.** [29]

Overview of upregulated enzymes and transporters (highlighted in red), involved in glycolysis ( hexokinase 2 (HK2), pyruvate dehydrogenase kinase 1 (PDK1), and lactate dehydrogenase A (LDHA)), lipid catabolism (hydroxyacyl-CoA dehydrogenase (HADH)), tricarboxylic acid (TCA) cycle and oxidative phosphorylation ( citrate synthase (CS), isocitrate dehydrogenase isoform 2 (IDH2), and alpha-ketoglutarate dehydrogenase-like (OGDHL)), serine and glycine pathways (e.g., phosphoglycerate dehydrogenase (PHGDH), phosphoserine aminotransferase 1 (PSAT1), and serine hydroxymethyltransferase 1/2 (SHMT1/2)), cysteine metabolism ( cystathionine-β synthase (CBS), cystathionase (CTH), and the cystine–glutamate antiporter (xCT), and glutamine metabolism (glutaminase 2 (GLS2), glutamate dehydrogenases (GLUD1/2), and glutamine transporter ASCT2).

# 4. Characterisation of the immune TME gives clues on immune responses and clinical outcomes in NB

NB patients exhibit distinct immune profiles, reflecting heterogeneity in immune responses and influencing patient outcomes. *In situ*, transcriptomic and single-cell analyses of the immune TME yielded conflicting results regarding the prognostic relevance of the different immune subsets in NBs.[44] High-risk NBs are classified as immunologically "cold" due to limited T-cell infiltration, downregulation of major histocompatibility complex class I (MHC-I) molecules which are essential for presenting tumour antigens to cytotoxic T cells[45], and low mutational burden, leading to an immunosuppressive tumour microenvironment (TME).[44] In particular, MycN-amplified tumors have reduced infiltration and activity of tumor-infiltrating lymphocytes compared to MycN-non amplified tumors, independent of tumor stage MYCN-A. NBs also showed reduced IFN pathway activity and lower expression of chemotactic and immune-activating mediatorsIFN pathway activity, promoted expression of CXCL9 and CXCL10 chemokines involved in T cell infiltration expression of activating NK ligands, resulting in increased NK cell cytotoxicity. Nevertheless, patients with T cell-inflamed HR tumors showed improved overall survival compared with those with non-T cell-inflamed tumors ($p<0.05$), independent of MYCN amplification status. In contrast, paraneoplastic neurological syndrome called Opsoclonus-Myoclonus Syndrome (OMS) or spontaneous regression observed in some LR NBs reflects the existence of a specific and effective antitumor immune response, which may explain the good oncological outcomes of such NBs.[1] A comparative study of the TME of 38 OMS NB patients and 26 non-OMS NB patients, including 13 HR and 13 LR, supports further the singularity of the OMS immune landscape.[46] OMS cases were characterized by TLS, a rich diversity of memory B and T cells, and elevated CD8+ T cells and activation markers (CTLA4, PD1) associated with high expression of genes such as CD22, BANK1, and TCF7. Half of OMS NB were classified as IFN-γ dominant subtypes, reminiscent of T cell activation. Furthermore, improved outcomes correlated with a diverse B cell IgH repertoire in OMS NB.[46] In contrast half of HR NB exhibited a TGF-β or wound-healing signature that is associated with poor prognosis. Another study has highlighted the key role of collaboration between natural killer (NK) and DCs (dendritic cells) in improving the survival of NB patients.[47]

Through the secretion of FLT3LG and CCL5, NK cells enable DCs recruitment within the TME. The presence of DC and NK cells correlated with T cell infiltrate. Furthermore, imaging analysis has revealed direct interaction of DCs and NK cells in tumour regions and near immune cell aggregates.[47]

Recently, several single-cell transcriptomic studies were performed on NB patient samples.[48–51] They brought major information about the immune and tumor heterogeneity in NBs. Bonine *et al* recently published a harmonized single-cell transcriptomic reference atlas of human NB tumors (NBAtlas) that integrates seven single-cell or single-nucleus including 362,991 cells across 61 patients.[48] These transcriptomic studies at single-cell resolution confirmed that immune cells have the ability to infiltrate even HR NBs. They also highlighted a differential repartition of immune cells according to survival and/or risk groups. The interplay between immune cells and malignant cell states has been reported.[6] Subsets and proportions of immune cells that infiltrate NBs were different from those present in the non tumoral adrenal gland. Myeloid cells, considered to be the population that preferentially infiltrates pediatric tumors, have been documented. Infiltration by conventional dendritic cells is associated with a better survival. A major challenge is to match the great heterogeneity of macrophages populations that infiltrate NB with a potential functional relevance. If specific macrophages activation states such as pro-inflammatory/anti-tumoral and immunosuppressive/pro-tumorigenic signatures are sometimes proposed as a prognostic indicator, the very rationale for such phenotypic assignment is challenged.[48] Still, an immunoregulatory network potentially mediated by myeloid cells have been highlighted in HR NB, affecting T and NK cell function.[49–51] Activated NK cells correlated with survival and chemotherapy rescue NK dysfunction.[49,51] Adaptive immune cells are also found in the NB immune landscape. Cytotoxic T cells, naive T cells, and Th17 signatures were significantly correlated with improved survival. Treg was not associated with survival, nor with exhaustion score in NB risk groups. Altogether these studies have deciphered the complexity of the TME in NB and allowed a more in depth comprehension of the cellular and molecular heterogeneity that drives tumour behavior and clinical outcomes.

# 5. New therapies aim to enhance antitumour immunity in NB

Disialoganglioside (GD2) was identified as a tumor-associated antigen, overexpressed in NB and has been used for diagnosis, also becoming a therapeutic target. Anti-GD2 antibodies like dinutuximab combined with irinotecan and temozolomide have improved objective response rates in relapsed/refractory NB up to 50–60%.[4] Similar results were observed with hu14.18K322A, a modified antibody that aims to promote NK and antibody-dependent cellular cytotoxicity activation.[4] Anti-GD2 coupled with anti-CD47 antibody (magrolimab) aiming to promote phagocytosis have shown synergistic effect and promising preclinical results. Despite the challenge of the TME bispecific antibodies like GD2 and CD3 to promote immune-mediated tumour cell killing, the purpose is to activate adaptive immunity through T cell activation. Anti GD2 CAR T cell therapies have also been used to promote transient immune activation and tumor response.[4] Another approach like invariant NK T cells and dual CAR constructs (e.g., GD2-triggered B7-H3 CARs) improve specificity by targeting multiple tumor-associated antigens and maintain effectiveness by reducing cell exhaustion.[4] Despite the improved outcomes offered by these therapies, 40% of patients relapse,[52] highlighting the diverse strategies tumours use for immune evasion. As we have seen, local immunity is influenced by local signals and metabolic conditions within the TME, as well as by systemic immunity.

The overall immune state can be influenced by chronic inflammation leading to a less effective immune response and tumour growth.[53] The relapses following GD2-targeted treatment and the heterogeneity of immune subtypes observed among each NB subtype highlight the need to consider that immunity differs across individuals and is shaped by each patient's unique history and characteristics.

# 6. Gut microbiome can influence tumour behavior and patient outcomes

The gut microbiome is a complex community of microorganisms, including bacteria, viruses, fungi, and other microbes, present in the gastrointestinal tract. It plays essential roles in many functions like digestion, immune system regulation, and the production of certain vitamins and metabolites.

Gut microbiome development begins in utero, influenced by microbial colonization from the mother via amniotic fluid swallowed by the fetus.[54] The diversity of the infant microbiome increases during birth, influenced significantly by exposure to maternal vaginal microbiota in normal deliveries compared to cesarean sections, which leads to increased risk of immune disorders like allergic rhinitis, asthma, and celiac disease.[55–57] After birth, the gut microbiome transitions through stages dominated first by Enterobacteriaceae and Staphylococcus, followed by Bifidobacterium and lactic acid bacteria during breastfeeding. After the introduction of solid foods, the microbiome diversifies further and stabilizes into an adult-like composition by approximately three years of age, undergoing significant changes in functionality to adapt to dietary shifts.[9]

Critical interactions between the host immune system and microbiota during early life have profound and lasting effects on immune homeostasis and susceptibility to diseases.[58] The gut microbiome plays a key role in shaping immunity since it engages in crosstalk communication with the immune system. Ligands of innate sensing receptors like TLRs (Toll-like receptors) or NLR (Nucleotide-binding oligomerization domain-like receptors) and metabolites like short-chain fatty acids (SCFAs) produced by the microbiome modulate gut immune cells and enterocytes.[58] These signals can also travel systemically, affecting global immunity. As examples, *Segmented Filamentous Bacteria* promotes Th17 differentiation while *Bacteroides fragilis* promotes CD4+ T cell differentiation and balances Th1/Th2 populations via its polysaccharide A (PSA), which is taken up by dendritic cells (DCs) and presented to naïve CD4+ T cells. In the presence of TGF-β, these cells become regulatory T cells (iTregs), producing IL-10 for immune homeostasis.[58] Treg and Th17 enhance secretory IgA's production thus maintaining a balanced microbiota. Altogether these data reinforce the concept that the gut microbiome can broadly impact host immune responses.

## 6.1 Gut microbiome alteration in NB's patients

Gut alteration was assessed in mice that received subperitoneal implantation of human NB cells.[59] After 10 weeks of tumour growth, gut microbiome analysis revealed alterations in specific bacterial species including trends of a decrease in *Firmicutes* and an increase in *Bacteroidetes*, *Deferribacteres*, and *Tenericutes*. In physiological conditions, *Firmicutes* dominate the gut microbiome after the first year of life producing SCFAs like butyrate.[9] Butyrate reduces inflammation by regulating cytokines and inducing Tregs. A decrease in *Firmicutes* may disrupt immune responses and increase gut inflammation.[9] *Bacteroidetes* promote nutrient absorption and produce SCFAs which serve as key energy sources and also modulate immune responses.[60] *Deferribacteres* play crucial roles in extreme environments through nutrient cycling, supporting gut ecosystem stability. Increased levels of *Bacteroidetes* and *Deferribacteres* in the context of NB may promote tumour cell proliferation by providing metabolic fuel and play a role in modulating immune responses, potentially influencing local and systemic inflammation associated with cancer progression.

A recent study also investigated the gut microbiome in NB patients and siblings compared to their mothers and healthy controls.[10] Microbiome analysis was performed considering several possible confounding factors such as age, sex, delivery mode, and dietary habits. NB patients exhibited reduced species richness in their gut microbiome compared to healthy controls, 18 species were found to be significantly lower among them *Phocaeicola dorei* and *Bifidobacterium bifidum* and butyrate-producers (*Roseburia*, *Faecalibacterium prausnitzii*). Only *Enterobacter hormaechei* was enriched in NB patients. Functional composition of the gut microbiome was altered in NB patients, with depletion in five metabolic pathways compared to controls, including starch biosynthesis from fructose, glycogen degradation to glucose 6-phosphate, synthesis of l-tyrosine and l-phenylalanine, and synthesis of vitamin B1 (thiamine diphosphate). NB patients showed reduced potential for carbohydrate degradation and aromatic amino acid synthesis. Additionally, they displayed significantly reduced saccharolytic fermentation and increased proteolytic fermentation compared to healthy controls, alongside decreased glycoprotein degradation potential. These alterations could create a metabolic environment favourable for NB progression.

It is well known nowadays that chemotherapies induce dysbiosis, significantly affecting homeostasis and overall metabolic processes. Cyclophosphamide is part of the standard chemotherapy of NB according to SIOPEN guidelines. The effect of this chemotherapy on the gut microbiome, inflammation and the gut barrier in a mouse model has been studied.[61] The results revealed that tumour-bearing mice that were treated by cyclophosphamide exhibited a pro-inflammatory state with reduced levels of *Lactobacillus* and had an increase in gut permeability that correlated with faecal volatile organic compound alteration.

## 6. 2 Causality and bidirectional relationships

Since NB pathogenesis has not been fully understood and is likely influenced by multiple factors it is legitimate to ask if gut microbiome disruption plays a role in it.

In a recent large-scale study, data from 18,340 individuals across 24 cohorts, primarily from Europe, and 4,881 NB samples sourced from the IEU Open GWAS Project were analyzed. Instrumental variables (IVs) were selected based on stringent criteria to ensure relevance and independence.[62] Mendelian randomization analysis identified six gut microbiota species as having a causal relationship to NB. *Lachnospiraceae* was a risk factor, its being positively correlated with the expression of brain-derived neurotrophic factor (BDNF), which, through the BDNF/TrkB pathway, promotes metastasis and invasion of NB cells via the PI3K/Akt/mTOR and MAPK pathways. In contrast, *Actinobacteria* exert a protective effect through the production of bioactive molecules, like antimycines, which inhibit complex III of the mitochondrial respiratory chain, thereby reducing the viability of NB cells and inhibiting their growth. *Bifidobacterium* inhibits the PI3K/Akt/mTOR pathway via galactose production and enhances immune responses through IL-27 expression. *Desulfovibrio* promotes NAC synthesis, leading to H2S production, which suppresses NB cell proliferation. *Howardella*'s protective role remains unexplored but may involve novel mechanisms.

A Genome-Wide Association Study data (ID ieu-a-816) with 1,627 NB cases and 3,254 control cases identified five significant associations.[63] Specifically, *Proteobacteria* showed a trend towards a protective effect while *Erysipelotrichachia displayed* a protective effect. More importantly, *Oscillospira intestinalis* was found to be linked to HR NB and *Erysipelotrichia* to LR NB. Gene expression analysis revealed that single nucleotide polymorphisms-related genes MUC4 and PELI2 were significantly correlated with several NB-associated genes. Pathway analysis revealed that MUC4 is involved in Wnt/Beta-catenin and KRAS signaling while PELI2 is linked to TGF-beta and p53 pathways. The findings were validated through cell line experiments, showing higher expression of MUC4 and PELI2 in aggressive MYCN-amplified NB cells. Altogether these data demonstrate that gut microbiota influences clinical outcome in NB through direct effects on tumor cells or indirect effect via immune modulation.

| Microbiome Alteration | Biological Consequences |
| --- | --- |
| Decrease in *Firmicutes* | Loss of butyrate production, promoting inflammation and disrupting immune regulation. |
| Increase in *Bacteroidetes* | Supports tumour cell metabolism through SCFAs and modulates immune responses, fostering inflammation. |
| Increase in *Deferribacteres* | Provides metabolic substrates, supporting tumor proliferation and immune modulation. |
| Reduced *Phocaeicola dorei*, *Bifidobacterium bifidum*, and butyrate-producers richness | Impaired gut homeostasis, reduced immune balance, and metabolic disruption. |
| Enrichment of *Enterobacter hormaechei* | Increased proteolysis weakens mucosal integrity and promotes inflammation. |
| Increase in *Lachnospiraceae* | Promotes tumor growth and metastasis |
| Increase in *Actinobacteria* | Produces bioactive molecules (e.g., antimycines) that inhibit mitochondrial function, reducing tumor cell viability and growth. |
| Decrease in *Bifidobacterium* | Impairs immune balance, inhibits galactose production, and reduces IL-27-mediated immune enhancement. |
| Increase in *Desulfovibrio* | Promotes NAC synthesis, leading to H2S production, which suppresses NB cell proliferation. |
| Increase in *Howardella* | Protective role likely involves novel mechanisms yet to be fully elucidated. |
| Increase in *Oscillospira intestinalis* | Linked to high-risk NB, potentially contributing to aggressive tumor behavior. |
| Increase in *Erysipelotrichia* | Associated with low-risk NB, possibly linked to more favorable outcomes. |

*Table* 2: **Microbiome alterations and their consequences in NB progression**

7. Targeting NB through dietary approaches

Nutrition emerges as a key strategy that can influence cancer cell metabolism, gut microbiome composition, and subsequently immune cell functions. For pediatric NB patients, nutritional interventions must support growth and development, help slow disease progression, manage diverse treatment side effects, and address unique patient and family needs.[64] The gut microbiome can easily be assessed with stool sample analysis, a non-invasive method that could be implemented in clinical routine due to its cost-effectiveness. Cancer cell metabolism can also be evaluated on blood or biopsy samples with mass spectrometry that can simultaneously analyze a wide range of metabolites, providing a comprehensive metabolic profile that can reflect the physiological state of the patient. In the following we discuss several findings that suggest that dietary interventions could have an impact on improving response to NB treatment or on overall survival (Figure 3).

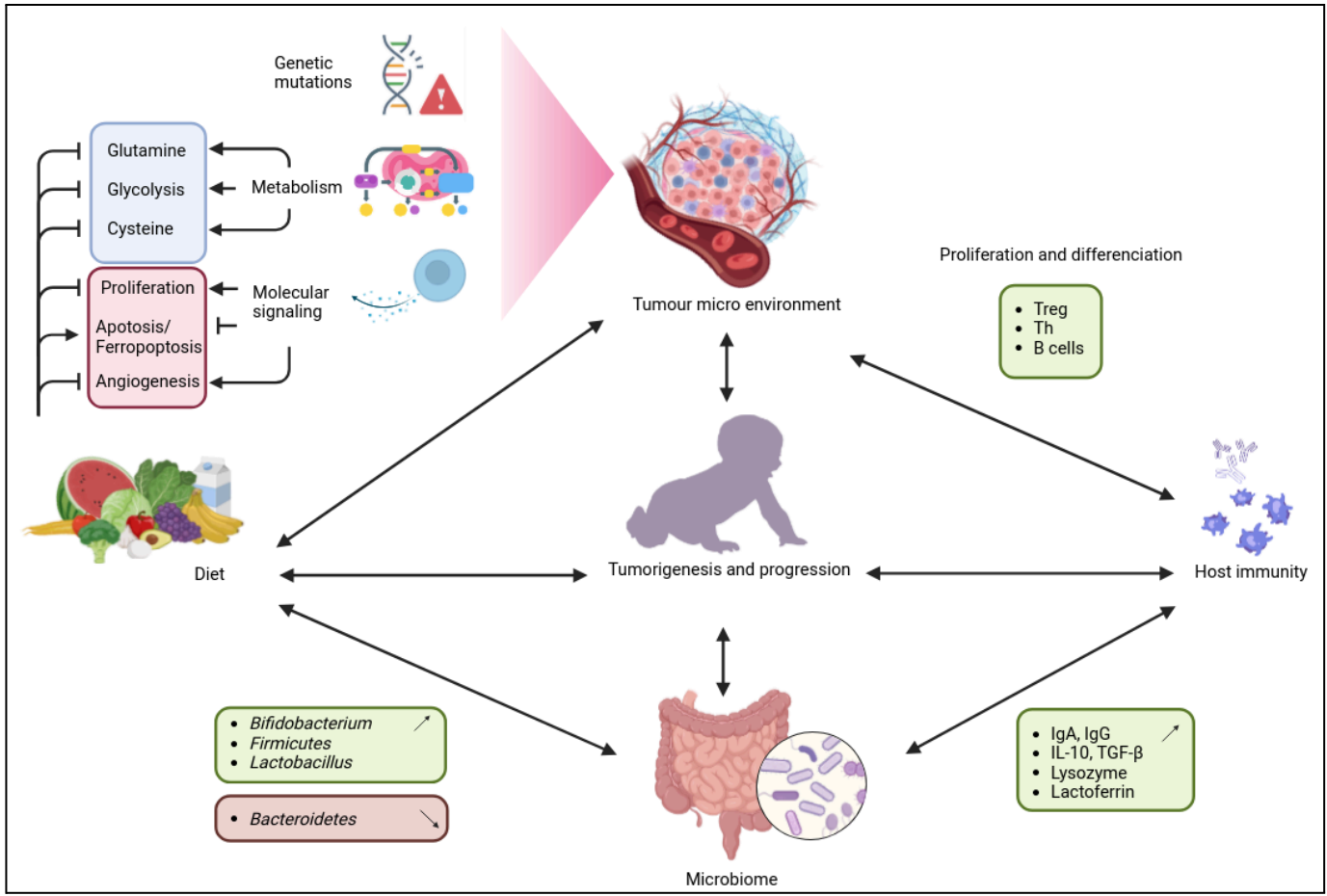

*Figure* **3**: **Personalised nutrition guidelines in neuroblastoma** [64]
Adapted illustration of path to personalised nutrition. Personalized nutrition in NB aims to restore gut microbiota balance, improving immune system modulation, and influencing tumor metabolism and progression in neuroblastoma patients. Proper nutrition supports the differentiation and proliferation of regulatory T cells (Tregs), helper T cells (Th cells), and B cells, enhancing immune surveillance and reducing chronic inflammation. Nutrients in the diet can influence the levels of immunoglobulins A and G (IgA, IgG), cytokines like Interleukin-10 (IL-10), Transforming Growth Factor-beta (TGF-β), lysozyme, and lactoferrin, which regulate immune responses and promote tolerance to commensal bacteria, creating a more favorable immunological environment. Additionally, tailored nutrition can impact the metabolic processes of the tumor, potentially slowing its growth and progression, and improving overall survival outcomes in neuroblastoma.

## 7.1 Impact of breastfeeding on infant health

Studies on several cohorts have highlighted the protective effect of breastfeeding and the use of folic acid and maternal use of any supplements containing folic acid, vitamins or minerals during the preconception period against NB.[11] Though less studied for NB than other cancers, breastfeeding may have a protective effect through several mechanisms.[9] Breast milk (BM) enables the transfer of beneficial bacteria like *Bifidobacterium* and *Lactobacillus*. This microbial transfer is crucial for establishing a healthy gut microbiota by reducing the risk of immune dysregulation and inflammation that can lead to cancer development. BM contains bioactive compounds such as human milk oligosaccharides that boost the growth of beneficial bacteria like *Bifidobacterium* that ferment human milk oligosaccharides into SCFAs. Some SCFAs like acetate, propionate or butyrate exhibit anti-inflammatory properties, regulating immune responses and promoting the proliferation and differentiation of Treg, Th and B cells, preventing chronic inflammation.[9] BM has immunomodulating properties since it exhibits essential immunomodulatory components, including immunoglobulins (IgA, IgG), cytokines (e.g., IL-10, TGF-β), lysozyme, and lactoferrin. These elements further regulate immune responses, promote tolerance to commensal bacteria, enhance immune surveillance and potentially lower the risk of tumorigenesis and of NB. Breastfed infants had lower levels of pro-inflammatory cytokines (IL-13 and IL-17A) and higher levels of anti-inflammatory cytokines (IL-10 and IFNβ). [9,65]

### 7.2 Ketogenic diet

The ketogenic diet (KD) is a high fat, adequate protein and low carbohydrate dietary therapy conventionally used to treat refractory epileptic children and that has shown potential as an adjuvant therapy for NB. [66,67] NB cells, and especially MYCN amplified ones, exhibit a high reliance on glycolysis and oxidative phosphorylation. It has been shown that KD induces a metabolic shift in NB cells reducing their ability to proliferate. Combining KD, with metformin, a drug that inhibits mitochondrial complex I and a low dose of cyclophosphamide, a chemotherapy agent, has demonstrated synergistic effects in reducing NB tumour growth.[67] Furthermore, KD has improved survival rates in MYCN amplified NB xenograft models by enhancing beta-oxidation. Another study has highlighted that KD is able to reduce NB tumour growth by inducing autophagy in mouse models.[68] KD creates a glucose-deficient TME that forces cancer cells to use fatty acids that are used less efficiently, thus rendering cancer cells more sensitive to chemotherapy or radiotherapy.[66,67] Moreover, KD negatively impacts angiogenesis through the downregulation of HIF-1α and vascular endothelial growth factor, thus contributing to the inhibition of tumour growth.[66] KD also reduces oxidative damage in healthy cells by reducing reactive oxygen species production while making tumour cells more vulnerable to cancer treatment.[66]

### 7.3 Cysteine deprivation as a trigger for ferroptosis

As we have mentioned earlier, cysteine is a key amino acid for energy production, and redox balance and is a source of carbon for NB cells, therefore its deprivation may hold potential benefits in NB nutritional strategy. Depletion of cysteine induces a reduction of the antioxidant glutathione that prevents lipid peroxidation that leads to ferroptosis. Cysteine deprivation in MYCN-amplified NB cells disrupts both extracellular uptake and the transsulfuration pathway, leading to oxidative damage and cell death.[69] This strategy may also synergize with ferroptosis inducers like glutathione peroxidase 4 inhibitors, offering a promising therapeutic approach to treating aggressive forms of neuroblastoma by exploiting their metabolic vulnerabilities.[69]

### 7.4 Glutamine

Glutamine deprivation is also one of the critical strategies that have emerged to target NB, particularly in MYCN-driven subtypes. Glutamine is essential for maintaining NB cell proliferation and survival.[40] Thus the deprivation of glutamine leads to the disruption of multiple cellular processes, including a reduction in MYCN and c-MYC expression, leading to decreased cell proliferation and viability.[70] Most surprisingly, glutamine deprivation combined with radiotherapy leads to selective radioresistance of MYCN amplified NB through compensatory mechanisms in MYCN proteins. This highlights the capacity of MYCN-amplified NB cells to adapt by increasing their cancer stem cell properties.[70] Furthermore, glutamine deprivation can have diverse effects on apoptosis depending on the chemotherapy used. It suppressed the etoposide-induced apoptosis by decreasing the expression of p53, which regulates death receptor 5 and caspase-8 activation.[71] In contrast, it enhances cisplatin-triggered cell death via ROS accumulation and mitochondrial dysfunction. Moreover, addition of 4-hydroxy-2-nonenal, a toxic lipid peroxidation product, to NB cells starved in glutamine medium enhances their mitochondrial membrane potential, leading to enhanced survival.[72] This potential was highly decreased by introducing etomoxir, which inhibits carnitine palmitoyltransferase 1. These findings collectively highlight how glutamine metabolism may be a key vulnerability in NB, although more research is needed on how to target this metabolic pathway.

### 7.5 Gut improvement strategies

Since NB patients display a decrease of *Firmicutes* and increase of *Bacteroidetes,* the modulation of gut microbiota through probiotic treatments could be a good strategy. In the treatment of inflammatory bowel disease, specific strains like *Lactobacillus reuteri* and *Lactobacillus plantarum* were shown to increase Firmicutes levels while simultaneously decreasing the abundance of Bacteroidetes. The therapeutic potential not only lies in restoring the balance between *Firmicutes*/*Bacteroidetes,* but also through their anti-inflammatory effect by reducing proinflammatory cytokines and promotion of beneficial microbial populations.[73] Increasing protective bacteria like *Bifidobacterium* can be supported by consuming fermented milk, yogurt or kimchi, depending on the age of NB patients.[74] A study investigating the use of prebiotics in a murine model of NB-induced tumor-associated cachexia, characterized by reduced muscle mass, showed promising results. Prebiotic treatment led to positive alterations in the gut microbiome, including an increase in beneficial bacteria (e.g., Clostridial Family XIII AD3011) and a decrease in harmful species (e.g., Muribaculum and Tyzzerella 3). These changes are likely to enhance the production of short-chain fatty acids (SCFAs), which may contribute to the observed effects. The study highlighted correlations between dietary intake, particularly carbohydrates, fibers, and vitamins and gut microbiome diversity, emphasizing the need for comprehensive dietary assessments in pediatric cancer care to potentially improve gut health and overall outcomes in NB patients.[75]

# 8. Conclusion

NB is a heterogeneous pediatric disease ranging from spontaneous regression to poor overall survival. The variability of NB might be attributed to the degree of differentiation of the initial tumour cell. The spectrum of phenotypes in NB is influenced by intrinsic factors like genetic alterations, chromosomal arrangements, and epigenetic modifications, which collectively contribute to disruption in homeostasis and a metabolic shift. Additionally, factors like gut alterations and host metabolic and immune status play crucial roles in tumours characteristics. Spontaneous regression that is observed in NB might be attributed to an efficient immune response, which we could aim to stimulate in worst prognosis cases. Stool analysis and mass spectrometry are non-invasive and cost-effective methods for clinical assessment of gut microbiome and tumour metabolic state. This information combined with the molecular profiling of the patients can inform clinicians about the tumour characteristics.

These insights could allow them to customize nutritional intervention in order to increase the treatment efficacy and improve the quality of life of NB's patients.

## Contributors

All authors participated in the conceptualisation and creation of the manuscript, literature search, writing, as well as original and final editing of the manuscript.

## Declaration of interests

We declare no competing interests

## Search strategy and selection criteria

References for this Review were identified through searches of Medline (PubMed) and Scispace database for Systematic Reviews for articles published up to August 1, 2024, using combinations of terms such as "Neuroblastoma", "epidemiology", "aetiology", "tumour micro environment", "molecular mechanism','metabolism", "treatment", "gut microbiome", "nutrition" and " diet". We also reviewed reference lists of published manuscripts, clinical guidelines, and other relevant reviews and meta-analyses.